\documentclass[
superscriptaddress,
nofootinbib,
nobibnotes,
notitlepage,
amsmath,amssymb,
aps,
twocolumn,
prl,
]{revtex4-2}

\usepackage{graphicx}
\usepackage[colorlinks]{hyperref}
\usepackage{bbding}
\usepackage{bm}
\usepackage{fontawesome5}
\usepackage{color}
\usepackage{xspace}
\usepackage{makecell}
\usepackage{multirow}
\usepackage{dcolumn}
\newcolumntype{d}{D{.}{.}{-1}}
\usepackage{booktabs}
\usepackage{threeparttable}
\usepackage[normalem]{ulem}

\def\beq{\begin{eqnarray}}
\def\eeq{\end{eqnarray}}

\usepackage[normalem]{ulem}
\usepackage{physics}
\usepackage{ifthen}
\usepackage{aas_macros}

\usepackage{xcolor}
\usepackage{pifont}

\begin{document}

\title{Dispu\texorpdfstring{$\bm{\tau}$}{tau}able: the high cost of a low optical depth}

\author{Noah Sailer{\textsuperscript{*$\dagger$}}}
\affiliation{Berkeley Center for Cosmological Physics, University of California, Berkeley, CA 94720, USA}
\affiliation{Lawrence Berkeley National Laboratory, One Cyclotron Road, Berkeley, CA 94720, USA}

\author{\,\,Gerrit~S. Farren{\textsuperscript{*$\ddag$}}}
\affiliation{Lawrence Berkeley National Laboratory, One Cyclotron Road, Berkeley, CA 94720, USA}
\affiliation{Berkeley Center for Cosmological Physics, University of California, Berkeley, CA 94720, USA}

\author{Simone Ferraro}
\affiliation{Lawrence Berkeley National Laboratory, One Cyclotron Road, Berkeley, CA 94720, USA}
\affiliation{Berkeley Center for Cosmological Physics, University of California, Berkeley, CA 94720, USA}

\author{Martin White}
\affiliation{Berkeley Center for Cosmological Physics, University of California, Berkeley, CA 94720, USA}
\affiliation{Lawrence Berkeley National Laboratory, One Cyclotron Road, Berkeley, CA 94720, USA}

\begingroup
\renewcommand{\thefootnote}{\fnsymbol{footnote}}
\footnotetext[1]{Both authors contributed equally to this work.}
\footnotetext[2]{\href{mailto:nsailer@berkeley.edu}{nsailer@berkeley.edu}}
\footnotetext[3]{\href{mailto:gfarren@lbl.gov}{gfarren@lbl.gov}}
\endgroup

\begin{abstract} 
Recent baryonic acoustic oscillation (BAO) measurements from the Dark Energy Spectroscopic Instrument (DESI) are mildly discrepant ($2.2\sigma$) with the cosmic microwave background (CMB) when interpreted within $\Lambda$CDM. When analyzing these data with extended cosmologies this inconsistency manifests as a $\simeq3\sigma$ preference for sub-minimal neutrino mass or evolving dark energy. 
It is known that the preference for sub-minimal neutrino mass from the suppression of structure growth could be alleviated by increasing the optical depth to reionization $\tau$.
We show that, because the CMB-inferred $\tau$ is negatively correlated with the matter fraction, a larger optical depth resolves a similar preference from geometric constraints.
Optical depths large enough to resolve the neutrino mass tension ($\tau\sim0.09)$ reduce the preference for evolving dark energy from $\simeq3\sigma$ to $\simeq1.5\sigma$ and increase the CMB-inferred values of $n_s$ and $H_0$ to $0.968\pm0.004$ and $67.94\pm0.44$ km/s/Mpc, respectively. 
Conversely, within $\Lambda$CDM the combination of DESI BAO, high-$\ell$ CMB and CMB lensing yields $\tau = 0.090 \pm 0.012$, which is in $\simeq3-5\sigma$ tension with \textit{Planck} low-$\ell$ polarization data when taken at face value.
Essentially all current CMB analyses $-$ including recent results from WMAP+ACT and SPT $-$ adopt the \textit{Planck} measurement of $\tau$: thus a systematic in large-scale
\textit{Planck} polarization would serve as a “single-point failure” for most modern cosmological analyses that include CMB data.
While there is no evidence for systematics in the large-scale \textit{Planck} data, $\tau$ remains the least well-constrained $\Lambda$CDM parameter and is far from its cosmic variance limit. This strengthens the case for future large-scale CMB experiments as well as direct probes of the epoch of reionization.
\end{abstract}

\maketitle

\section{Introduction}
\label{sec:introduction}

Recent measurements of the cosmic microwave background (CMB) and baryon acoustic oscillations (BAO) have highlighted a number of discrepancies within the standard flat $\Lambda$CDM cosmological model. For example, DESI DR2 BAO together with CMB measurements from \textit{Planck} show a $\sim 3 \sigma$ preference for a time-varying (dynamical) dark energy model \cite{DESIDR2}. These stem in part from parameter inconsistencies (such as in the matter density $\Omega_{\rm m}$ and the Hubble parameter $H_0$) in the comparison between the values inferred at early times by the CMB and at later times by BAO within $\Lambda$CDM.

Moreover, attempts at measuring the mass of neutrinos through their effect on cosmological observables \cite{Lesgourgues:2006nd, Loverde:2024nfi} have revealed surprising features: when imposing a positivity prior, the constraints are tighter than a Fisher information analysis would suggest. If a positivity prior is not imposed, and the ``effective'' neutrino mass is allowed to be negative, the minimum mass implied by oscillation experiments ($\sum m_{\nu} \geq 0.06$ eV \cite{NuFit:2024eli}) is excluded at $\sim 3 \sigma$ \cite{DESI:2025ejh, Elbers:2024sha, Green:2024xbb, Craig:2024tky, Lynch:2025ine}. These discrepancies can be partially alleviated by e.g. dynamical dark energy \cite{DESI:2025ejh} or new physics in the neutrino sector \cite{Craig:2024tky}. However, given the paradigm shift this would imply, it is important to explore other possibilities. 

It has long been known that most measurements of neutrino mass from the suppression of structure growth are sensitive to the assumed value of the optical depth to reionization $\tau$ \cite{2015PhRvD..92l3535A,Yu:2018tem,2024PhRvD.109j3519G, Loverde:2024nfi, Craig:2024tky,2025arXiv250305691A}. Moreover, constraints on $\tau$ are also correlated with other cosmological parameters (e.g. $\Omega_{\rm m}$ and $H_0$) and therefore $\tau$ influences constraints derived from the expansion history, including both neutrino mass and the time-evolution of dark energy \cite{2024PhRvD.109j3519G,2025arXiv250305691A}.

Typically measured through large-scale polarization of the CMB at $\ell \lesssim 20$, the last ``official'' \textit{Planck} analysis gives $\tau = 0.0544 \pm 0.0073$ \cite{Planck:2018vyg} based on 2018 PR3 data. Subsequent analyses of the same data find a slightly higher value $\tau \approx 0.06$ \cite{deBelsunce:2021mec, Pagano:2019tci, BeyondPlanck:2020rud}. Independent analyses of \textit{Planck} PR4 data yield $\tau = 0.0533 \pm 0.0074$ \cite{Rosenberg:2022sdy} and $\tau = 0.058 \pm 0.006$ \cite{Tristram:2023haj}. While great progress in the understanding and control of systematics has been made in the past few years, this remains an extremely challenging measurement. As noted in the \textit{Planck} legacy paper \cite{Planck:2018nkj}, the polarization on large scales is more than two orders of magnitude smaller than the corresponding temperature anisotropies, making it difficult to control instrumental systematic effects and foregrounds while retaining enough sensitivity to the signal of interest (see Section 6.6 of \cite{Planck:2018nkj} or further discussions in refs~\cite{Pagano:2019tci,2022MNRAS.517.2855D,2020A&A...643A..42P,Tristram21,Rosenberg:2022sdy,2024A&A...682A..37T}).

Moreover, the interpretation of the polarization signal depends to some extent on the reionization history: $\tau$ is almost independent of the detailed history for ``fast'' models of reionization \cite{Planck:2016mks}, but a non-trivial fractional ionization at high redshift could potentially affect the CMB measurement of $\tau$ \cite{Heinrich:2021ufa}. We do not attempt to study shifts in $\tau$ given different reionization histories, but we note that no preference for a high redshift ionized component is apparent in the latest \textit{Planck} data, and that an ionized component at $z>15$ appears to be disfavored by the large-scale CMB polarization \cite{Millea:2018bko, Planck:2018vyg}, although this statement is somewhat model dependent \cite{2025arXiv250413254I}.

Direct measurements of the Lyman-$\alpha$ forest and other observables tightly constrain the low-$z$ half of reionization \cite{McGreer:2014qwa, Greig:2016vpu, Greig:2018rts, Sobacchi:2015gpa, Mason:2019ixe, Whitler:2019nul, Wang:2020zae} (and are consistent with a ``late and fast'' reionization); however, the higher-$z$ tail has very few direct constraints. Observations of high redshift galaxies from the James Webb Space Telescope (JWST) are consistent with a non-zero ionization fraction at $z \gtrsim 10$ (but with significant uncertainty), potentially leading to a larger optical depth \cite{Munoz:2024fas, 2024ApJ...967...28N, 2024ApJ...975..208T}. New physics, such as dark matter annihilation or decay \cite{Liu:2016cnk}, or accretion onto primordial black holes \cite{Belotsky:2017vsr, Ricotti:2007au}, could also increase the ionization fraction at even higher redshift, potentially affecting CMB measurements of $\tau$.

Other techniques, such as using the redshifted 21cm line from neutral hydrogen \cite{Liu:2015txa,Sailer:2022vqx} or the kinematic Sunyaev-Zel'dovich (kSZ) effect \cite{2013ApJ...776...83B, Park:2013mv, Jain:2023jpy, Nikolic:2023wea, Smith:2016lnt, Ferraro:2018izc, Alvarez:2020gvl}, might shed light on the matter in the future. The absence of a detection of a large kSZ power spectrum or trispectrum appears to favor a relatively ``short'' reionization, though this is model dependent and currently relatively weak \cite{SPT:2020psp, SPT-3G:2024lko, MacCrann:2024ahs}.

To explore the sensitivity of the tensions discussed here to $\tau$, we exclude low-$\ell$ polarization \cite{2020A&A...641A...5P,2019A&A...629A..38D,2024A&A...682A..37T,deBelsunce:2021mec} from our fits, and replace it with a Gaussian prior on $\tau$. We study two scenarios: one with prior $\tau = 0.06 \pm 0.006$, which reproduces the \textit{Planck} constraints very closely, as well as a hypothetical higher value $\tau = 0.09 \pm 0.006$. 
This latter prior is, at face value, inconsistent with the \textit{Planck} low-$\ell$ measurement at $3.8\sigma$ significance \cite{note:prior}.
Such a large shift is likely not reconciled by changes in the reionization history alone; however, given the considerations outlined above, we think that it is interesting to explore the consequences of a hypothetical higher value of $\tau$.

Following previous work \cite{Planck:2018vyg, Yu:2018tem, 2024PhRvD.109j3519G,Loverde:2024nfi, Craig:2024tky}, we also consider the inverse problem: assuming a flat $\Lambda$CDM model and a physical prior on neutrino masses from oscillation experiments, we solve for $\tau$ (without the inclusion of the low-$\ell$ polarization), and show that the inferred $\tau$ is in tension with the \textit{Planck} measurement, thus suggesting that within flat $\Lambda$CDM, some of the internal inconsistencies can be reinterpreted as a ``$\tau$ tension.''

Given that modeling errors could potentially be at play we choose to only include observables that can be modeled using linear theory (primary CMB, BAO) or with extremely mild non-linearities and no additional free parameters (CMB lensing). The inclusion of other observables such as galaxy clustering, galaxy weak lensing, cross-correlations, or supernovae is worthwhile but requires care and will be the subject of future work.  

\section{Data}
\label{sec:data}

We adopt the shorthand \textbf{``DESI"} for the DR2 BAO data \cite{DESIDR2} and \textbf{``CMB"} for the combination of \textit{Planck} PR4 \verb|CamSpec| high-$\ell$ \cite{Efstathiou:2019mdh,Rosenberg:2022sdy} primary temperature and polarization measurements with PR4 \cite{Carron:2022eyg} and ACT DR6 \cite{ACT:2023dou,ACT:2023ubw,ACT:2023kun} lensing data.
We substitute low-$\ell$ CMB data \cite{note:noTT} with Gaussian priors on the optical depth $\tau$ with mean 0.06 or 0.09 and uncertainty $\sigma_\tau=0.006$, corresponding to the approximate $\sigma_\tau$ from low-$\ell$ \textit{Planck} data \cite{Planck:2018vyg}. We use the shorthand \textbf{``$\bm{\tau=0.06}$"} and \textbf{``$\bm{\tau=0.09}$"} for these priors respectively.
To directly compare with previous results in the literature we choose not to include the ACT DR6 primary CMB likelihood \cite{ACT:2025fju, ACT:2025tim, ACT:2025xdm}, while noting excellent agreement between ACT and \textit{Planck} on the relevant parameters and the absence of a $\tau$ measurement from ACT.

\section{The optical depth's impact on cosmological `tensions'} \label{sec:tau_higher}

\begin{figure}[!h]
    \centering
    \includegraphics[width=0.39\linewidth]{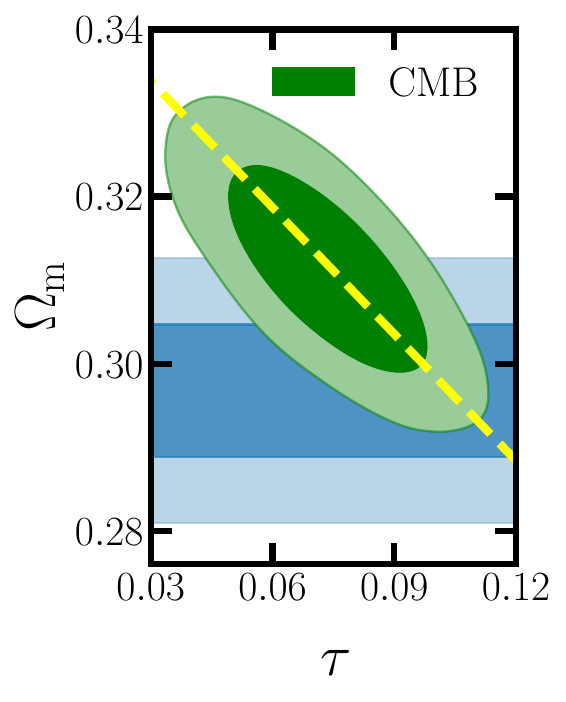}
    \includegraphics[width=0.595\linewidth]{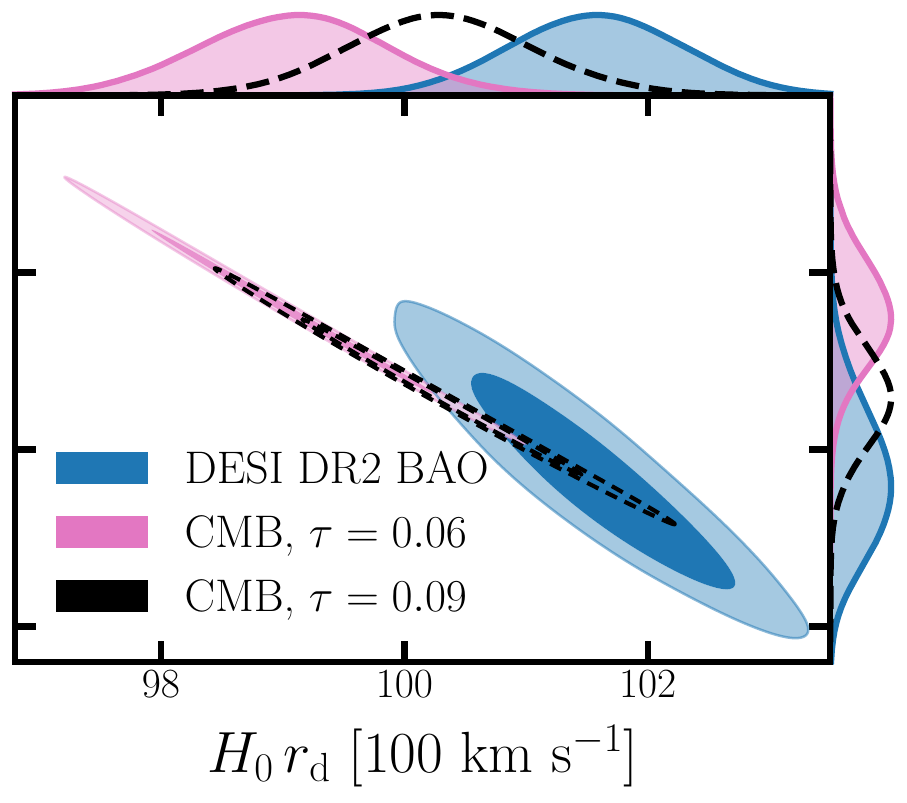}
    \caption{
    \textit{Left:} Constraints on $\tau$ and $\Omega_{\rm m}$ from the joint analysis of high-$\ell$ CMB and CMB lensing. The dashed yellow line $\big(\Omega_{\rm m}(\tau)=0.349-0.505\,\tau\big)$ approximates the $\tau-\Omega_{\rm m}$ degeneracy direction.
    \textit{Right:} $\Lambda$CDM constraints from BAO (blue) or CMB high-$\ell$ primary and lensing with a $\tau$ prior (pink/black).
    }
    \label{fig:tau_H0rd_OmM_contours}
\end{figure}

When jointly fitting to CMB lensing and high-$\ell$ primary CMB the inferred optical depth and matter fraction $(\Omega_{\rm m})$ are negatively correlated within $\Lambda$CDM, as illustrated in the left panel of Fig.~\ref{fig:tau_H0rd_OmM_contours}.
Thus increasing the optical depth brings the CMB-preferred $\Omega_{\rm m}$ into better agreement with BAO measurements.
We illustrate this in the right panel of Fig.~\ref{fig:tau_H0rd_OmM_contours}, where we plot constraints on $\Omega_{\rm m}$ and the product of $H_0$ with the sound horizon at the drag epoch ($r_{\rm d}$) from DESI BAO data (blue)
alongside those derived from the CMB when additionally including a $\tau=0.06$ (pink) or $\tau=0.09$ (black) prior. 
Following section VI of ref.~\cite{DESIDR2} we compute the parameter-level $\chi^2$ using the best fit $(\Omega_{\rm m},\,H_0r_d)$ values from DESI and the CMB, finding $\chi^2 = 6.55$ and $2.07$ when including a $\tau=0.06$ or $0.09$ prior respectively. The corresponding $\Lambda$CDM ``tension" between the BAO and CMB data reduces from $2.1\sigma$ to $0.9\sigma$. 

\subsection{Neutrino mass}
\label{sec:neutrino_mass}

The combination of primary CMB, CMB lensing and BAO has shown a preference for neutrino masses in tension with lower bounds from oscillation experiments \cite{Elbers:2024sha, Craig:2024tky, Green:2024xbb,Loverde:2024nfi}. 
To assess the consistency of the inferred neutrino mass sum with the lower bound $\sum m_\nu \geq 0.06$ eV we implement a phenomenological model in which the neutrino mass is replaced by an effective parameter $M_{\nu, \rm{eff}}$. While there is not a unique way to extend the model to negative values of $M_{\nu, \rm{eff}}$ \cite{Elbers:2024sha, Craig:2024tky, Green:2024xbb}, here we use the following simple prescription: For any observable $X(\Sigma\,m_\nu)$, we take
\begin{equation}
    X(M_{\nu, \rm{eff}}) = X(0)+ {\rm sgn}(M_{\nu, \rm{eff}})\Big[X(\vert M_{\nu, \rm{eff}}\vert) - X(0)\Big].
\end{equation}
This implementation recovers the physical neutrino model for $M_{\nu, \rm{eff}}>0$ and linearly extrapolates to negative ``masses." 
This model was shown to agree well with a more sophisticated treatment based on replacing the neutrino energy density with an effective energy density, allowed to take on negative values, at all orders in perturbation theory \cite{Elbers:2024sha}. We are able to reproduce the results based on the latter method in \cite{DESI:2025ejh} with good accuracy.
When including large-scale polarization data from \textit{Planck} we find $M_{\nu,\mathrm{eff}} = -0.105^{+0.049}_{-0.062}$ (cf. $-0.101^{+0.047}_{-0.056}$ \cite{DESI:2025ejh}).

\begin{figure}[!h]
    \centering
    \includegraphics[width=0.9\linewidth]{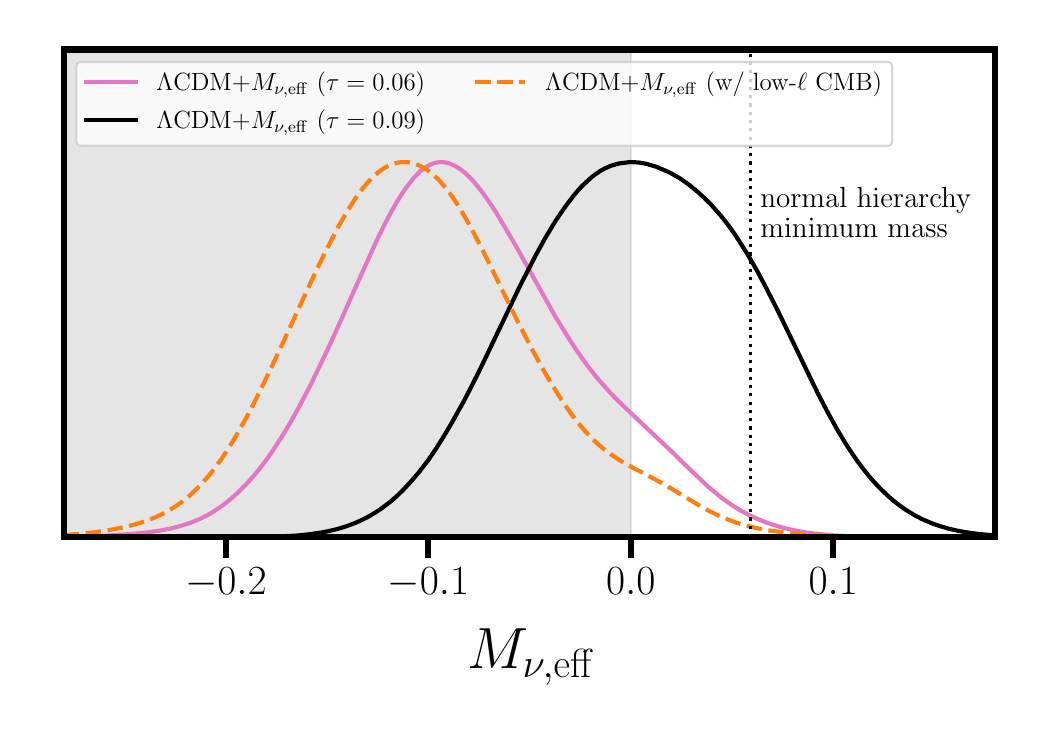}
    \caption{Constraints on $M_{\nu,\mathrm{eff}}$ from the joint analysis of BAO and CMB data when including a $\tau$ prior  ($\tau=0.06$ in pink and $\tau=0.09$ in black) or low-$\ell$ data from \textit{Planck} (orange).
    }
    \label{fig:neutrino_mass}
\end{figure}

When replacing the low-$\ell$ \textit{Planck} data with Gaussian priors centered on $\tau=0.06$ and $\tau=0.09$ we find $M_{\nu,\mathrm{eff}} = -0.08^{+0.05}_{-0.06}$ and $M_{\nu,\mathrm{eff}} = 0.01\pm 0.06$ respectively (see Fig.\,\ref{fig:neutrino_mass}). While the former exhibits a moderate $\sim 2.4\sigma$ tension with the minimum mass allowed in the normal hierarchy, the latter is consistent with minimum mass, normal hierarchy neutrinos at $\sim 0.9\sigma$. Both values of $\tau$ provide similarly good fits to the data with a $\Delta \chi^2_{\rm MAP}=1.8$ in favor of the smaller of the two.

Our results are consistent with refs.~\cite{Craig:2024tky,Loverde:2024nfi}, who note that for $\tau\sim0.09$ the resulting increase in $A_s$ shifts the neutrino mass inferred from the suppression of structure towards its minimal value. 
The negative correlation between $\tau$ and $\Omega_{\rm m}$ (Fig.~\ref{fig:tau_H0rd_OmM_contours}) supports this trend since the ``geometric"  preference for negative neutrino mass weakens as the CMB-inferred $\Omega_{\rm m}$ decreases.

Note that other ways of alleviating the tension between $\Omega_{\rm m}$ inferred from the CMB and BAO, like the decaying dark matter model presented in \cite{Lynch:2025ine}, do not alleviate the discrepancy in both methods of inferring the neutrino mass, since they increase the amplitude of matter density fluctuations inferred from CMB lensing and therefore potentially exacerbate any excess lensing compared to $\Lambda$CDM expectations.

\subsection{Evolving dark energy}
\label{sec:ede}

\begin{figure}[!h]
    \centering
    \includegraphics[width=0.9\linewidth]{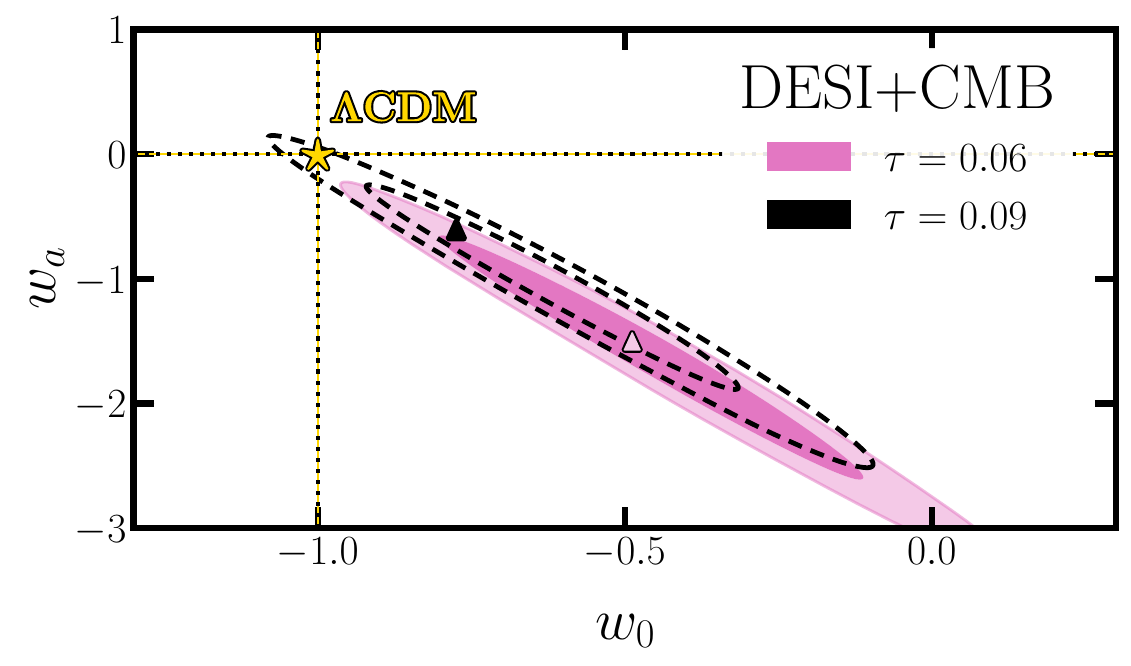}
    \caption{
    Constraints on the $w_0w_a$ model from the combination of high-$\ell$ primary CMB data, CMB lensing, DESI BAO and a prior on the optical depth.
    Locations of posterior maxima are indicated by triangles.}
    \label{fig:w0_wa_contours}
\end{figure}

The DESI collaboration \cite{DESIDR2} finds a $3.1\sigma$ preference for the $w_0w_a$ dynamical dark energy model \cite{Chevallier:2000qy, Linder:2002et} over $\Lambda$CDM based on a joint analysis of BAO and CMB (primary and lensing) measurements.
When substituting a $\tau=0.06$ prior in place of the \textit{Planck} low-$\ell$ data we find $\Delta \chi^2_{\rm MAP}=-10.75$ between the best-fit $w_0w_a$CDM and $\Lambda$CDM cosmologies. This corresponds to a $2.8\sigma$ preference for the former, in reasonable agreement with the DESI DR2 results. 
As shown in Fig.~\ref{fig:w0_wa_contours}, the best fit $(w_0,w_a)$ shifts from $(-0.5,-1.5)$ to $(-0.8,-0.6)$ when increasing the mean of the $\tau$ prior from 0.06 to 0.09. 
With $\tau=0.09$ we find $\Delta\chi^2_{\rm MAP}=-3.82$, corresponding to a mild, $1.4\sigma$, preference for evolving dark energy. 

\section{Inferring the optical depth without reference to large-scale CMB}\label{sec:infer_tau}

\begin{figure}[!h]
    \centering
    \includegraphics[width=0.9\linewidth]{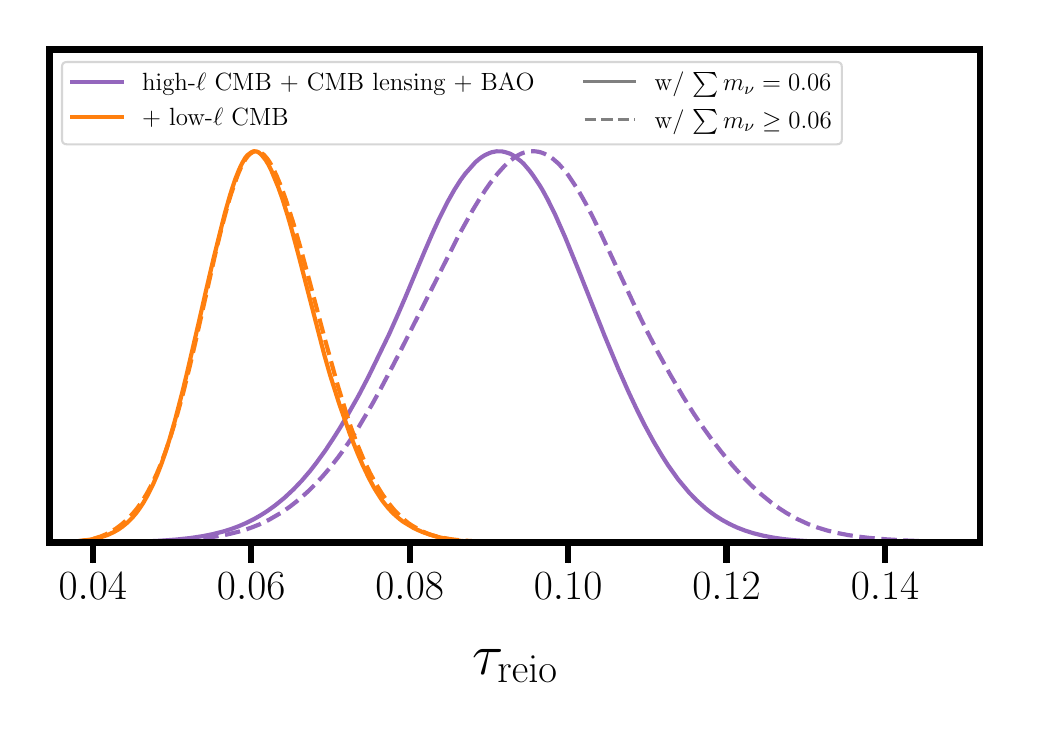}
    \caption{When discarding the low-$\ell$ data from \textit{Planck} we infer $\tau=0.090\pm 0.012$ within $\Lambda$CDM from the combination of high-$\ell$ CMB, CMB lensing and BAO, or $\tau=0.095\pm 0.014$ for $\Lambda{\rm CDM}+\sum m_\nu$ with a physical prior of $\sum m_\nu \geq 0.06$ eV. 
    }
    \label{fig:tau_free}
\end{figure}

We have shown that several widely discussed preferences for beyond-$\Lambda$CDM physics arising from CMB and BAO data are substantially weakened when a larger optical depth is assumed. 
Conversely, we may ask the question what value of $\tau$ is preferred by the data when combined within a given model and whether this value is consistent with the one inferred from low-$\ell$ CMB polarization.
For this purpose we combine the high-$\ell$ CMB, CMB lensing, and BAO data, but leave $\tau$ to vary with a uniform prior in the range $0.02$ to $0.20$. We consider two scenarios, one in which the neutrino mass sum is fixed to its minimal mass ($\sum m_\nu = 0.06$ eV) and one where it may to take on any value allowed by oscillation constraints ($\sum m_\nu \geq 0.06$ eV).
We find 
\begin{equation}
\begin{aligned}
&\tau = 0.090 \pm 0.012\,\,\,\,(\Lambda{\rm CDM})\\
&\tau = 0.095 \pm 0.014\,\,\,\,(\Lambda{\rm CDM}+\Sigma m_\nu),\\
\end{aligned}
\end{equation}
as illustrated in Fig.~\ref{fig:tau_free}. 
These values are in $\sim 3\sigma$ disagreement with the optical depth inferred from the full CMB data set (including large-scale polarization) in combination with CMB lensing and BAO \cite{note:correlation}, and are consistent with those found by refs.~\cite{2024PhRvD.109j3519G,Loverde:2024nfi}.

\section{Discussion and Conclusions}
We have explored the cosmological consequences of a hypothetical higher value of $\tau$ and shown that some of the recently discussed inconsistencies between CMB and BAO data when interpreted within $\Lambda$CDM can be alleviated by a higher value of $\tau$. 
These findings build upon previous results showing that a larger $\tau$ reduces tensions internal to the \textit{Planck} data \cite{2024PhRvD.109j3519G}.

We summarize our main findings in Table~\ref{tab:consistent_with} and further note that with a $\tau=0.09$ prior the CMB-inferred values of $n_s$ and $H_0$ increase to $0.968\pm0.004$ and $67.94\pm0.44$ km/s/Mpc within $\Lambda$CDM \cite[see also][]{2024PhRvD.109j3519G,2025arXiv250305691A}.
We have not included data from supernovae, galaxy weak lensing, galaxy clustering or cross-correlations. We note that type Ia supernovae are themselves in mild discrepancy with $\Lambda$CDM \cite{Brout:2022vxf, Rubin:2023ovl, DES:2024jxu}; however, their joint analysis with the CMB is also affected by the discussion in this Letter.

\begin{table}[!h]
    \centering
    \begin{tabular}{|c|c|c|}
    \hline
        Consistent with? & $\tau = 0.06$  & $\tau = 0.09$  \\ \hline
        \hline
        \textit{Planck} low-$\ell$ polarization & \CheckmarkBold & \XSolidBrush \XSolidBrush \,\cite{note:caveat} \\ \hline
        WMAP low-$\ell$ polarization & \CheckmarkBold & \CheckmarkBold \\ \hline
        BAO within $\Lambda$CDM & \XSolidBrush & \CheckmarkBold \\
        \hline
        $\sum m_{\nu} \geq 0.06$ eV & \XSolidBrush & \CheckmarkBold \\ \hline
        Cosmological constant $\Lambda $& \XSolidBrush & \CheckmarkBold \\ \hline
    \end{tabular}
    \caption{Summary of our findings. Here we define \XSolidBrush as ``inconsistent'' at $>2 \sigma$, and \CheckmarkBold as ``consistent'' within $2\sigma$. 
    }
    \label{tab:consistent_with}
\end{table}

We caution that at present we have no evidence to doubt the CMB polarization measurements of $\tau$. However, given the difficulties of the measurement and the paradigm shift in terms of fundamental physics that some of the proposed solutions would imply, we think that it is important to explore all possibilities. 
We emphasize that since current CMB analyses $-$ including recent results from WMAP+ACT \cite{ACT:2025fju, ACT:2025tim, ACT:2025xdm} and SPT \cite{SPT-3G:2025bzu} $-$ rely on the \textit{Planck} $\tau$ measurement, a systematic in large-scale \textit{Planck} polarization would serve as a ``single-point failure" for essentially all modern cosmological analyses that include CMB data.
A combination of a statistical fluctuation and mild systematics in one or more of the datasets might also be able to reconcile the results. In fact, $\tau$ values as low as 0.08 can bring all of the discrepancies discussed below $2\sigma$.

Finally, we have highlighted the crucial role that $\tau$ plays not only for neutrino masses, but also for exploring new physics with cosmological observations.
Given that current observational uncertainties remain well above the cosmic variance limit \cite{Watts:2019uvq}, we believe that there is a strong scientific motivation for future experiments designed to achieve higher precision, as well as for more direct measurements of the epoch of reionization by observing high redshift galaxies, the 21cm signal or the kSZ effect, among others.

Scripts used to organize this analysis are publicly available on the \verb|disputauble| \href{https://github.com/NoahSailer/disputauble}{\faGithub} GitHub repository \cite{note:github}.
This Letter makes use of the \verb|matplotlib| \cite{4160265},  \verb|numpy| \cite{Harris2020-hc} and \verb|scipy| \cite{Virtanen2020-rh} packages. We use the Boltzmann code \verb|CAMB| \cite{Lewis_2000,Howlett_2012} to compute theory spectra, and use \verb|GetDist| \cite{2019arXiv191013970L} and \verb|Cobaya| \cite{Torrado_2021} for inference.

\section*{Acknowledgments}

We thank Liang Dai, Roger de Belsunce, George Efstathiou, Lloyd Knox, Arthur Kosowsky, Marilena Loverde, Gabriel Lynch, Matthew McQuinn, Seshadri Nadathur, Alexander Reeves, Uro\v{s} Seljak, Blake Sherwin and Zachary Weiner for useful discussions.
NS additionally thanks Mr\href{https://noahsailer.github.io/assets/images/catmath.jpg}{.} May.
NS, GSF, and SF are supported by Lawrence Berkeley National
Laboratory and the Director, Office of Science, Office
of High Energy Physics of the U.S. Department of Energy under Contract No. DE-AC02-05CH11231.

\subsection*{Data availability}

The data that support the findings of
this article are openly available \cite{note:zenodo}.

\bibliography{main.bib}

\begin{thebibliography}{86}
\bibitem [{not({\natexlab{b}})}]{note:noTT}%
  \BibitemOpen
  \bibinfo {note} {We additionally discard low-$\ell$ temperature for
  simplicity. We have explicitly checked that CMB-derived $\Lambda$CDM
  constraints shift by at most $0.2\sigma$ with the inclusion of low-$\ell$
  temperature.}\BibitemShut {Stop}%
\bibitem [{not({\natexlab{d}})}]{note:caveat}%
  \BibitemOpen
  \bibinfo {note} {At face value, the inconsistency is at $\sim 5 \sigma$,
  reduced to $\sim 3.8 \sigma$ when accounting for the prior width. However, we
  note the caveats discussed in the introduction.}\BibitemShut {Stop}%
\end{thebibliography}

\end{document}